# Unveiling the Impact of Crosslinking Redox-Active Polymers on Their Electrochemical Behavior by 3D Imaging and Statistical Microstructure Analysis

Marten Ademmer [a,§], Po-Hua Su [b,c,§], Lukas Dodell [a], Jakob Asenbauer [b,c], Markus Osenberg [d], André Hilger [d], Jeng-Kuei Chang [e,f], Ingo Manke [d], Matthias Neumann [a,*], Volker Schmidt [a], Dominic Bresser [b,c,*]

[a] *Institute of Stochastics, Ulm University, 89069 Ulm, Germany*

[b] *Helmholtz Institute Ulm (HIU), 89081 Ulm, Germany*

[c] *Karlsruhe Institute of Technology (KIT), P.O. Box 3640, 76021 Karlsruhe, Germany*

[d] *Institute of Applied Materials, Helmholtz-Zentrum für Materialien und Energie, 14109 Berlin, Germany*

[e] *Department of Materials Science and Engineering, National Yang Ming Chiao Tung University, Hsinchu 30010, Taiwan*

[f] *Department of Chemical Engineering, Chung Yuan Christian University, Taoyuan 32023, Taiwan*

***Corresponding authors:** dominic.bresser@kit.edu; matthias.neumann@uni-ulm.de

[§]These authors contributed equally.

**Abstract**

Polymer-based batteries offer potentially higher power densities and a smaller ecological footprint compared to state-of-the-art lithium-ion batteries comprising inorganic active materials. However, in order to benefit from this potential advantages, further research to find suitable material compositions is required. In the present paper, we compare two different electrode composites of poly(2,2,6,6-tetramethylpiperidinyloxy-4-ylmethacrylate) (PTMA) and CMK-8, one produced with and one without crosslinking the PTMA. The influence of both approaches on the corresponding electrodes is comparatively investigated using electrochemical measurements and statistical 3D microstructure analysis based on synchrotron X-ray tomography. A particular focus is put on the local heterogeneity in the coating and how the crosslinking influences the interaction between PTMA and CMK-8. It is shown that crosslinked PTMA—compared to its non-crosslinked counterpart—exhibits a more heterogeneous microstructure and, furthermore, leads to better surface coverage of CMK-8, larger pores and shorter transportation pathways through the latter. These changes improve the electrochemical properties of the electrode.

## 1. Introduction

Lithium-ion batteries (LIBs) are the dominating battery technology for small-scale applications such as portable electronics and large-scale applications including (hybrid) electric vehicles.[1,2] Nevertheless, important components such as graphite, the active material for the negative electrode, as well as nickel and cobalt, comprised in $LiNi_{1-x-y}Mn_xCo_yO_2$ as the active material for the positive electrode, are considered critical concerning their long-term supply—especially in Europe.[3-5] To address these potential issues, alternative battery chemistries are needed that benefit from the use of more abundant elements and components. In fact, as not all applications eventually need the very high energy density provided by LIBs, sodium-ion batteries, for instance, are considered a viable (complementary) alternative.[6-8] Another alternative candidate is given by organic batteries, relying mostly on carbon—ideally derived from biomass—as an essentially unlimited resource.[9-11] One of the most studied organic active materials for the positive electrode is poly(2,2,6,6-tetramethylpiperidinyloxy-4-yl methacrylate) (PTMA) owing its relatively high dis-/charge potential of about 3.6 V vs. $Li/Li^+$ and excellent rate capability.[12-15] However, the dissolution of PTMA in organic electrolytes and the resulting continuous capacity loss remained an issue, hindering its practical application.[16-18] One strategy to overcome this issue relies on crosslinking the PTMA in order to decrease the solubility.[16,18-21] As a result, the PTMA-based electrodes showed higher capacities and substantially improved cycling stability compared to the non-crosslinked analogues.[16,18,20,21] What remained unexplained somehow, though, is the substantially higher capacity recorded for the crosslinked PTMA despite the reduced radical concentration, the slower charge transfer kinetics, and the reduced swelling with the electrolyte.[18] These findings suggest that other factors play an important role for the achievable capacity—presumably factors that are dependent on the active material, i.e., the PTMA, itself. One such factor might be the accessibility of the redox active moieties—or in other words the spatial distribution of PTMA in the electrode and its overall microstructure. An appropriate methodology to quantitatively study the 3D electrode microstructure is the combination of tomographic imaging with image analysis, using methods of spatial statistics and mathematical morphology for investigating the resulting image data.[22, 23] This allows for the computation of morphological microstructure descriptors which are, on the one hand, experimentally not accessible like the lengths of shortest transportation paths and, on the other hand, have a strong impact on effective physical properties such as effective ionic diffusivity.[24, 25] In previous papers, this methodology has been widely used to study the morphology of differently manufactured electrodes in LIBs with $LiNi_{1-x-y}Mn_xCo_yO_2$ as the cathode active material,[26-29] where the use of machine learning opens new possibilities for an appropriate segmentation of image data, i.e., to reliably reconstruct the 3D microstructure from grayscale images. Recently, the influence of different binder materials on the morphology of classical PTMA-based electrodes has been studied for the first time,[30] where synchrotron X-ray tomography (XT) has been the method of choice to resolve the morphology of the 3D microstructure at the electrode scale.[31, 32]

In the present paper, we use a combination of XT and statistical image analysis to investigate the impact of crosslinking PTMA on the electrochemical behavior. Following an optimization of the electrode composition by replacing part of the commonly used conductive carbon nanoparticles by a nanostructured mesoporous carbon we conducted statistical image analysis based on a machine learning supported segmentation of image data, performed with the aid of the software Ilastik.[33] The results show that the crosslinking has a beneficial impact on the 3D electrode microstructure, leading to a pore network with improved connectivity properties and a larger interface between pores and solid material without significantly compromising the length of shortest transportation paths through the solid phase.

## 2. Materials Synthesis, Experimental Details & Basic Characterization

*Chemicals and Materials*

The following reagents and solvents for the polymerization were used as received without additional purification: 4-methacryloyloxy-2,2,6,6-tetramethylpiperidine-*N*-oxyl (TEMPO-M, 98%, Sigma-Aldrich), 1-methoxy-2-methyl-1-trimethylsilyloxy-propene (MTS, 97%, Alfa Aesar), tetra-n-butylammonium fluoride (TBAF, 1M in tetrahydrofuran (THF), Alfa Aesar), THF (water-free, max. 0.003% $H_2O$, >99.9%, stabilized with Ionol, VWR), and ethylene glycol dimethacrylate (stabilized with hydroquinone monomethyl ether for synthesis, Sigma-Aldrich).

*PTMA Synthesis and Crosslinking*

The PTMA was synthesized via a group transfer polymerization as reported by Bugnon et al.[16] and depicted in **Figure S1**. The synthesis was performed in an argon-filled glove box with an $O_2$ and $H_2O$ content of less than 0.1 ppm. In brief, for the synthesis of the non-crosslinked PTMA (G-PTMA), 4.0 g TEMPO-M were dissolved in 9.6 mL THF under continuous stirring. After 5 min, 124 µL MTS and 46 µL TBAF were added dropwise into the solution. The polymerization was allowed to proceeded for 40 h in the glove box. Subsequently, the 0.8 mL methanol were added and the resulting mixture was stirred for 5 min, before pouring it dropwise into 100 mL hexane provided in a round-bottom flask. The slightly pink, essentially white precipitate was collected as the final product. The remaining liquid phase was further stirred for 12 h to maximize the product yield by collecting the eventually remaining orange gel-like precipitate. The merged solid phases were dried at 60 °C under vacuum (< $10^{-2}$ mbar) for 12 h. The overall yield was about 80%. For the synthesis of the crosslinked PTMA (XG-PTMA), 0.0329 g ethylene glycol dimethacrylate were added at the very beginning to the 4.0 g TEMPO-M dissolved in 9.6 mL THF. All further steps remained the same.

*Physicochemical Characterization*

The successful synthesis of G-PTMA and crosslinked XG-PTMA was evaluated by means of Fourier-transform infrared (FTIR) spectroscopy (PerkinElmer Spectrum Two, attenuated total reflectance mode) and differential scanning calorimetry (DSC, Discovery series, TA Instruments). (**Figure S2**). The FTIR spectrum of XG-PTMA showed an additional absorption band at 1634 $cm^{-1}$, corresponding to ethylene glycol dimethacrylate that was used as crosslinker (**Figure S2a**), indicating the successful crosslinking. This was further corroborated by the DSC data (**Figure S2b**), revealing a shift of the glass transition temperature from 156 °C to 168 °C owing to the increased rigidity and decreased dynamics. Moreover, G-PTMA showed a melting temperature $T_m$ at 98 °C, which was absent for XG-PTMA. Scanning electron microscopy (SEM) and energy-dispersive X-ray spectroscopy (EDX) were performed using a ZEISS EVO MA 10 electron microscope equipped with an EDX detector (Oxford Instruments X-MaxN, 50 $mm^2$), applying an acceleration voltage of 3 kV (SEM) and 10 kV (EDX). Nitrogen sorption measurements were carried out using a Quadrasorb SI (Quantachrome), and the specific surface area and pore size distribution of the CMK-8 were determined by the Brunauer-Emmett-Teller (BET) and Barrett-Joyner-Halenda (BJH) method, respectively. Prior to these measurements, the samples were outgassed at 120 °C to remove potentially adsorbed water from the surface.

*Electrode Preparation*

A first batch of electrodes was prepared comprising only nanoparticulate conductive carbon (SuperC65, SC65, Imerys) as the electron-conducting additive. The electrode active material was either G-PTMA or XG-PTMA, and the electrodes are referred to as G-PTMA-SC65 and XG-PTMA-SC65, respectively. For the preparation of these electrodes, 54 wt% (X)G-PTMA, 41 wt% SC65, and 5 wt% sodium carboxymethyl cellulose (CMC, WALOCEL$^{TM}$ CRT, Dow Wolff Cellulosics) were mixed by manual grinding before adding deionized water. The resulting slurry was stirred for 12 h, before casting it with a laboratory-scale doctor blade on an aluminum foil serving as the current collector (wet film thickness: 120 µm). The thus prepared electrodes were initially dried at 80 ºC for 6 h, before punching disc-shaped electrodes with a diameter of 12 mm. These electrodes were further dried under vacuum ($<10^{-2}$ mbar) at room temperature for 3 h and at 120 °C for 12 h.

For the second batch of electrodes, the conductive additive SC65 was partially replaced by CMK-8 (ACS Material). Initially, 60 mg (X)G-PTMA were manually mixed with 40 mg CMK-8 in 0.5 mL 2-butoxyethyl acetate (BCA) to enhance the contact between the two components. To fully dissolve the PTMA, the temperature of the mixture was increased to 80 ºC and maintained at this temperature for 4 h, before further increasing it to 160 °C and maintaining it at this further elevated temperature for 12 h to completely evaporate the BCA. The thus obtained powder was additionally dried at 80 °C under vacuum ($<10^{-2}$ mbar) overnight. The subsequent electrode preparation was the same as for the first batch of electrodes, except the resulting difference concerning the electrode composition, which was 54:36:5:5 (X(G)-PTMA / CMK-8 / SC65 / CMC) by weight. The resulting electrodes were referred to as XG-PTMA-CMK-8 and G-PTMA-CMK-8.

*Cell Assembly and Electrochemical Characterization*

Three-electrode Swagelok-type cells were assembled in an argon-filled glove box ($O_2$ < 0.1 ppm and $H_2O$ < 0.1 ppm). Lithium metal foil (Honjo) served as counter and reference electrode and the electrodes were separated by glass microfiber sheets (Whatman, GF/A). The separator between the working and counter electrode was drenched with 120 µL of the electrolyte, i.e., 1M $LiPF_6$ in a 1:1 volume mixture of ethylene carbonate (EC) and dimethyl carbonate (DMC), purchased from UBE. The cells were subjected to galvanostatic cycling using a Maccor 4000 battery tester. The testing temperature was kept constant at 20 °C by placing the cells in a climatic chamber (Binder). The cut-off voltages were set to 3.0 V and 4.0 V vs. $Li/Li^+$. Prior to the cycling, the cells were allowed to rest for 12 h. The given values for the specific current, ranging from 25 mA $g^{-1}$ to 50 A $g^{-1}$, refer to the mass of the active material.

## 3. 3D Microstructure Characterization – Tomographic Imaging & Statistical Data Analysis

*Synchrotron X-ray Tomography*

To further characterize and compare the electrodes, prepared as described in Section 2, in terms of their morphology at the micro-scale, three-dimensional (3D) images were captured using XT. The synchrotron tomography experiments were carried out at the P05 beamline at PETRA III (DESY, Germany).[34] The samples were exposed to a synchrotron energy of 20 keV. During the tomography, the sample was continuously rotated over 180 degrees, while 3001 projections were acquired with an exposure time of 0.3 s each. The sample rotation axis had a distance of 15 mm to the $CdWO_4$ scintillator screen, resulting in a small amount of phase contrast. The scintillator projection image was magnified with a 10 times microscope optic and then captured using a KIT CMOS camera yielding a pixel size of 0.64 µm.

For the reconstruction of the tomography data, the filtered back-projection based software developed at DESY was used.[35] After reconstruction, an additional total variation denoising filter was applied.[36] The resulting 3D image data are the basis for the following microstructure characterization.

*Image Segmentation Using Random Forests*

Before the microstructure could be characterized by statistical analysis of the 3D image data, representing the electrode microstructure, some preprocessing steps were required. First, the 3D images are rotated to align them with the coordinate system of the sampling window. Next, a global threshold and a morphological opening using a sphere of radius 5 voxel as structuring element were applied to roughly separate the aluminum foil of each sample from the solid phase, pores and background. Here and in the following, solid phase means the solid phase of the coating. Based on this preliminary foil segmentation, the polynomial approach described by Westhoff et al.[37] was used to straighten the sample, i.e., to align the aluminum foil such that it is parallel to the *x*-*y*-plane. Note that a final segmentation of the aluminum foil was obtained by means of random forests, as described below. After this alignment, additional steps were performed to determine a mask, in which the electrode was contained, and subsequently, to create a segmentation separating solid phase, pore space and aluminum foil. As shown in **Figure 1a**, there were large variations in the grayscale values in that part of the image, which we would visually attribute to the solid phase. Due to this variation it was not possible to classify a voxel either as solid phase, pore space or aluminum merely based on its grayscale value. In the following, an appropriate three-step procedure is proposed to get an accurate segmentation of the data. In short, the Sobel edge detection operator[38] in conjunction with a cluster analysis was used to create a mask of the electrode, i.e., to roughly separate the electrode structure from the surrounding background. From this, a morphological closing was used to determine the boundaries of the sample and, finally, a random forest classifier was used to create the final segmentation. To determine the mask, the Sobel edge detection operator was applied to the aligned grayscale image data, creating gradient images. Those images were binarized by global thresholding to identify the interfaces between pores and solid. The threshold was chosen in such a way that most of the edges form a connected phase in 3D. A cluster analysis was used to find all connected components in the complement of this edge phase. The two largest clusters were the two regions above the coating and below the aluminum foil, i.e., the background of the sample. To remove artifacts resulting from the edge detection, a morphological closing with a sphere of radius 11 voxel as structuring element was applied to the background. The mask, which contains the electrode structure, was finally given by the complement of the background. A 2D example slice of the mask for the sample without crosslinking is shown in **Figure 1b**. In the next step, the aluminum foil, the solid phase and the pore space had to be separated from one each other within the mask. As mentioned before, simply assigning each voxel to one of the three phases using global thresholds was not appropriate. Therefore, two random forest classifiers[39] were trained using Ilastik[33] based on hand-labeled training data. First, a random forest was trained, which separated the aluminum foil from its complement. Subsequently, a second random forest was trained classifying all voxels, not already determined as aluminum foil, as either pores or solid. This approach was convenient, as the number of hand-labeled areas needed to create an accurate segmentation was reasonably small. An example of a hand labeled 2D slice is shown in **Figure S3**. Thus, we obtained 3D image data, where the voxels were classified as aluminum foil, solid phase or void, i.e., pores classified by the random forest or voxels outside the mask. To distinguish between inner pores, which are not necessarily identical with the pores determined by the random forest, we used the rolling ball algorithm with a radius of 20 voxels as described by Charry et al.[40]. In **Figure 1c**, an example of the resulting segmentation is shown, on the basis of which the microstructure analysis described in the next section was performed.

*Microstructure Characterization by Statistical Image Analysis*

The separation of solid phase, pore space and aluminum foil in the 3D image segmentation as described in the previous section enabled us to characterize the microstructure of the electrodes by means of statistical image analysis. For this purpose, global as well as local microstructure descriptors and pairwise interdependencies between them are discussed for each sample. Methodologically similar microstructure characterizations are presented in[30,41] for differently manufactured PTMA-based electrodes and paper-based materials, respectively.

To begin with, the microstructure descriptors for both samples had to be computed. A uniform hexahedral grid of points, with a distance of 50 µm between neighboring grid points, was superimposed over each sample in *x*-*y*-direction, i.e., the plane parallel to the electrode coating and aluminum foil. Each point was the center of a sampling window for the computation of local microstructure descriptors. These non-overlapping sampling windows extended over the complete sample thickness, or, more precisely, from the aluminum foil to the opposite boundary in *z*-direction. The edges of these sampling windows in *x*-*y*-direction were increased from 10 µm to 50 µm in 10 µm steps. This led to 189 sampling windows across the electrode based on G-PTMA-CMK-8 and 330 across the XG-PTMA-CMK-8 electrode. In **Figure 2**, a 3D rendering of a cutout with 100 µm is shown to give a visual impression of the sample differences and local variability of the samples themselves. Note that the computed descriptors of such a sampling window are the averaged results computed for the respective local sampling window. Additionally, note that for the descriptors considered in the present paper, the mean computed across all sampling windows with an edge length of 50 µm is equal to the mean of the complete sample. These mean values are global microstructure descriptors and are summarized in **Table 1**. As descriptors, we considered the volume fraction of the solid phase, the electrode thickness, the surface area per unit volume (SAV) of the solid phase, as well as the mean geodesic tortuosities of the solid phase and the pore space. For a given point in the *x*-*y*-plane, its pointwise thickness was defined as the distance between the first and the last voxel (with the same *x*-*y*-coordinates) of the electrode in *z*-direction. The thickness of a predefined sampling window was then given as the average value of all pointwise thicknesses for points within the sampling window under consideration. To compute the surface area from voxelized image data, we used the method described previously.[42] Mean geodesic tortuosity quantifies the length of transportation paths[24, 43]. It was defined as the mean length of shortest transportation paths starting from the aluminum foil intersected with this cutout and reaching to the opposite boundary of the electrode through the considered material phase. Note that these paths were allowed to leave the given local sampling window to avoid edge effects.

**Table 1.** Global descriptors quantifying the microstructure of the G-PTMA-CMK-8 and XG-PTMA-CMK-8 electrodes.

| | **Volume fraction solid phase / %** | **Thickness / µm** | **Surface area per unit volume / µm$^{-1}$** | **Mean geodesic tortuosity (pore space)** | **Mean geodesic tortuosity (solid phase)** |
|---|---|---|---|---|---|
| **G-PTMA-CMK-8** | 76 | 22.4 | 0.274 | 1.24 | 1.03 |
| **XG-PTMA-CMK-8** | 65 | 26.7 | 0.294 | 1.11 | 1.05 |

Considering the global microstructure descriptors in **Table 1**, we observed that the cross-linked sample was thicker and exhibited a smaller volume fraction of the solid phase, i.e., a higher porosity. The increased porosity is most likely the reason for the smaller mean geodesic tortuosity of the pore space in the cross-linked case, while the mean geodesic tortuosity of the solid phase was close to 1 in both cases. Moreover, the specific surface area per unit volume was 7% larger for XG-PTMA-CMK-8 compared to G-PTMA-CMK-8. Note that, similar to the observations in[30], there seems to be a negative correlation between volume fraction and SAV of the solid phase for volume fractions above 60%, since the XG-PTMA-CMK-8 sample had a slightly higher global SAV compared to the G-PTMA-CMK-8 electrode (0.294 $\mu m^{-1}$ vs. 0.274 $\mu m^{-1}$). A larger SAV can be advantageous, as it quantifies the amount of boundary at which the electrolyte and the solid phase can interact. We will come back to this relationship when discussing the interdependence of local microstructure characteristics below in detail.

Besides the global morphology of electrode microstructures, we also quantitatively investigated local heterogeneities. **Figure 2** shows the heterogeneity of both samples at the microscale, while G-PTMA-CMK-8 seemed to be more homogeneous than XG-PTMA-CMK-8. In particular, comparing these cutouts, larger variations in thickness and volume fraction of the solid phase were observed for the coating of the XG-PTMA-CMK-8 sample. These visual impressions were confirmed by our statistical analysis (see the distributions of local volume fraction of the solid phase in **Figure 3a,e**). Even though the G-PTMA-CMK-8 sample exhibited relatively strong heterogeneities, where variations of volume fraction of the solid phase were especially pronounced for the smaller sampling windows, the XG-PTMA-CMK-8 sample was considerably more heterogeneous. Especially for sampling windows with an edge length of 10 µm the variations of volume fraction of the solid phase were considerably larger. While some sampling windows had a volume fraction below 40%, it was above 90% in other sampling windows. As expected, this gap became smaller for increasing sizes of the sampling windows. The distributions of local SAV, visualized in **Figure S4**, show a similar behavior as the distributions of local volume fractions, where a larger variability was observed for XG-PTMA-CMK-8.

Similar observations were made with regard to the distributions of local thickness of both samples shown in **Figure 3b,f**. The decrease of variability for an increasing window size was much less pronounced compared to that of local volume fractions for both samples. The distributions of local thickness were considerably wider across all sizes of sampling windows for XG-PTMA-CMK-8. Even for sampling windows with an edge length of 50 µm, the local thickness of XG-PTMA-CMK-8 varied more than the local thickness for sampling windows of G-PTMA-CMK-8 with an edge length of 10 µm. An interesting observation, however, was that while the local thickness distributions of XG-PTMA-CMK-8 were nearly symmetrical, they were clearly positively skewed for G-PTMA-CMK-8. Another important observation was the difference in mean thickness between both samples. Since the wet films on the aluminum foil exhibited the same thicknesses before drying, the difference in terms of thickness was apparently a result of the drying process. Furthermore, as both material compositions consisted of the same weight percentage of solid materials, we conclude that the drying created larger pores in the coating of the XG-PTMA-CMK-8 sample. This matches the visual impression and is in good accordance with the larger volume fraction of the solid phase observed for the G-PTMA-CMK-8 electrode.

The local distributions of mean geodesic tortuosities of the pore space and the solid phase are shown in **Figures 3c,g** and **3d,h**, respectively. Due to the lower values of local porosity of G-PTMA-CMK-8, it is not surprising that the mean geodesic tortuosity of the pore phase of this electrode was larger and varied more than that of the XG-PTMA-CMK-8 electrode. For the mean geodesic tortuosity of the solid phase, the opposite was observed. The generally higher volume fraction of the solid phase in the G-PTMA-CMK-8 sample led to shorter transportation paths in the solid phase, thereby decreasing the mean geodesic tortuosity.

For the considered electrode materials, a more interesting quantity is the mean geodesic tortuosity of the pore phase, since the supply of electrons in the solid phase is rarely a limiting factor regarding the performance of the cell. XG-PTMA-CMK-8 not only had a significantly lower mean value with 1.11 compared to the 1.24 of G-PTMA-CMK-8, but additionally, just a few of the considered local sampling windows had a mean geodesic tortuosity exceeding 1.2. On the other hand, some windows even reached 1.5 in the case of G-PTMA-CMK-8. It is important to note that these differences were especially significant in this context, as the samples were rather thin in comparison to the typical cluster size in the solid phase. In other words, since many clusters of the solid phase exhibited the same thickness as the complete coating, the amount of obstructions decreased significantly at a certain distance from the start, leading to many paths being nearly straight after the first few voxels. Therefore, it is to be expected that the length of the shortest paths would increase more slowly with the sample thickness, decreasing the mean geodesic tortuosity of both samples and additionally widening the gap between them.

On the other hand, since many of these clusters directly connect the foil with the opposite boundary, the values of mean geodesic tortuosity of the solid phase were nearly 1 for both samples. Even in the extreme cases, in which the mean geodesic tortuosity through the solid phase reached values of up to 1.1 for the XG-PTMA-CMK-8 electrode, these values were still smaller than the global mean geodesic tortuosity of the pore phase.

Finally, to quantify the interdependence between pairs of local microstructure descriptors, we considered the corresponding bivariate distributions. More precisely, we considered the bivariate distributions of local porosity and each of the other local microstructure descriptors considered in the present paper.

In general, the larger heterogeneity of XG-PTMA-CMK-8 was also reflected in these bivariate distributions. However, the only clear difference between the samples with respect to bivariate distributions was that there was a stronger relationship between thickness and volume fraction of the solid phase for XG-PTMA-CMK-8 compared to G-PTMA-CMK-8, see **Figure 4a,e**. The other bivariate distributions showed similar trends in both cases. Nevertheless, an interesting observation was made with regard to the bivariate distribution of the volume fraction and mean geodesic tortuosity of the pore phase. More precisely, there did not seem to be a correlation between both microstructure descriptors, as seen in **Figure 4b,f**. This was true for both samples and was especially surprising taking into account that, as discussed before, a smaller volume fraction in general leads to shorter transportation paths in the pore space. Furthermore, **Figure 4c,g** shows that, as expected, the volume fraction and mean geodesic tortuosity of the solid phase were clearly negatively correlated. This different behavior of shortest paths in the pore space and the solid phase can be explained by considering two opposite influences of local thickness on the local mean geodesic tortuosity. In particular, in the solid phase, the shortest transportation paths often circumvent the first obstacle, i.e., the first cluster of the solid phase they meet, and go then directly to the separator. Just considering this effect would lead to a negative correlation between local thickness and local mean geodesic tortuosity. On the other hand, the effect that thickness positively correlates with porosity would lead to a positive correlation between local thickness and local mean geodesic tortuosity of the solid phase. Our results, in which no correlation was visible, indicate that these two effects were balanced for the electrodes considered in this paper.

Finally, note that in **Figure 4d,h** a negative correlation between surface area per unit volume and volume fraction of the solid phase, also mentioned when discussing the global microstructure descriptors, was observed. Moreover, the mean values of SAV conditioned on the local volume fraction showed a maximum at approximately 60%. This coincides with the findings reported by Neumann et al.[30].

## 4. Electrochemical Characterization and Discussion

First, we investigated the electrochemical behavior of G-PTMA and XG-PTMA in a standard electrode configuration using SC65 as the electron conductive additive. The results are presented in **Figure S5**, revealing a superior specific capacity for the crosslinked XG-PTMA-SC65 electrodes – in line with previous papers on the impact of crosslinking.[16,18,20,21] However, the capacity remained generally rather low with about 60 (XG-PTMA-SC65) and 50 mAh $g^{-1}$ (G-PTMA-SC65), respectively. Thus, we replaced part of the nanoparticulate SC65 with nanostructured mesoporous CMK-8, following previous studies on such material as conductive additive, which showed a highly beneficial impact of such carbon.[44-46] The basic characterization of CMK-8 is provided in **Figure S6**, revealing a particle size of several micrometers with frequently more than 10 µm (**Figure S6a**), a specific surface area of 574 $m^2$ $g^{-1}$ (**Figure S6b**), and a mean pore size of around 22 Å (**Figure S6c**). SEM micrographs of the resulting electrodes, i.e., G-PTMA-CMK-8 and XG-PTMA-CMK-8, are displayed in **Figure 5a** and **Figure 5b**, respectively. Generally, the G-PTMA-CMK-8 electrodes showed less of the relatively large holes in the electrode coating layer, while the XG-PTMA-CMK-8 electrode morphology resembled largely the morphology of the CMK-8 particles (cf. **Figure S6a,b**). This indicates that crosslinking the PTMA yielded a better surface coverage of the CMK-8 conductive carbon, which was confirmed by the EDX mapping presented in **Figure 5c-f** and, in particular, the EDX mapping of nitrogen in **Figure 5e,f**. In fact, in the case of G-PTMA-CMK-8 (**Figure 5e**), the PTMA appeared to be well distributed across the overall electrode, while it was only detectable on the larger "edgy" particles in the case of XG-PTMA-CMK-8 (**Figure 5f**). The more intimate contact with the CMK-8 might be beneficial for the performance of the electrodes following previous studies reporting the need for an intimate mixing of PTMA and nanosized carbons.[20, 21]

To investigate this, the G-PTMA-CMK-8 and XG-PTMA-CMK-8 electrodes were subjected to galvanostatic cycling with a lithium-metal counter electrode (**Figure 6**). While both electrodes showed increased capacities compared to the CMK-8-free G-PTMA-SC65 and XG-PTMA-SC65 electrodes (**Figure S5**) with 54 (G-PTMA-CMK-8) and 88 mAh $g^{-1}$ (XG-PTMA-CMK-8), the capacity increase was more pronounced for XG-PTMA-CMK-8 (**Figure 6a**). In the latter case, also the capacity retention was much greater with 89% compared to only 65% for the non-crosslinked PTMA after 100 cycles, which underlines the beneficial impact of crosslinking concerning the solubility in the liquid organic electrolyte.[19] The rate capability of the two different electrodes was evaluated as well and the results are presented in **Figure 6b**. In general, XG-PTMA-CMK-8 provided higher specific capacities at all specific currents with, e.g., 91 vs. 71 mAh $g^{-1}$ at 25 mA $g^{-1}$ and 52 vs. 23 mAh $g^{-1}$ at 1 A $g^{-1}$. Nonetheless, the gap between the two different electrodes narrowed at elevated specific currents, suggesting that especially at very high specific currents of several A $g^{-1}$ the positive impact of the crosslinking vanishes. The comparison of the dis-/charge profiles (**Figure 6c-f**) revealed that for G-PTMA-CMK-8 (**Figure 6c,e**) there is essentially only a shortening of the voltage plateau, while for XG-PTMA-CMK-8 (**Figure 6d,f**) the same behavior was only observed for a specific current up to 0.1 A $g^{-1}$. When the current was further increased, additionally a significant increase in polarization was observed – still yielding a higher specific capacity, though. In fact, the more pronounced increase in polarization is even more evident from the comparison of further increased specific currents (**Figure S7a-d**), resulting in a greater voltage hysteresis compared to G-PTMA-CMK-8 (**Figure S7e,f**). This observation is in line with the preferential CMK-8 surface coverage of the crosslinked PTMA, leading to a higher polarization owing to its electronically insulating nature.

The combination of 3D imaging with statistical microstructure analysis further elucidated these differences between G-PTMA-CMK-8 and XG-PTMA-CMK-8 in terms of their electrochemical behavior. Differences with regard to the spatial distribution of pores and PTMA were also visible in the 3D image data and could be quantified by the selected microstructure descriptors. In particular, the statistical analysis of image data revealed that the XG-PTMA-CMK-8 electrode was considerably more heterogeneous. This can be attributed to the fact that the PTMA in the XG-PTMA-CMK-8 electrode seems to better cover the CMK-8 leading to a higher number of large clusters of the solid phase. Moreover, it is shown that the local volume fraction of the solid phase and the local sample thickness positively correlate (**Figure 4a,e**), meaning that the local thickness of the coating is highly dependent on the presence of clusters of the solid phase. Compared to XG-PTMA-CMK-8, the electrode comprising G-PTMA-CMK-8 was thinner on average and the solid phase was more homogeneously distributed with only a relatively small number of large clusters. Thus, the sample with XG-PTMA-CMK-8 exhibited not only a higher porosity, but also larger pores. This results in considerably different behavior of the considered samples regarding mean geodesic tortuosity quantifying the length of shortest transportation paths. First, the transportation paths through the pore phase of the XG-PTMA-CMK-8 electrode were significantly shorter compared to those of G-PTMA-CMK-8. Second, the surface area per unit volume was larger for XG-PTMA-CMK-8 than for G-PTMA-CMK-8, i.e., XG-PTMA-CMK-8 had a larger interface at which the electrolyte in the pores can interact with the solid phase and, thus, with the PTMA. Both properties are beneficial for the dis-/charge performance of the electrode.

Moreover, it is reasonable to presume that this larger SAV amplified the beneficial effect of the improved surface coverage of CMK-8 by the crosslinked PTMA. Note that, even though one could expect the mean geodesic tortuosity of the solid phase to get worse with crosslinked PTMA due to the higher porosity, only a negligibly small increase of mean geodesic tortuosity was observed.

These changes of the 3D microstructure are presumably one of the factors leading to another observation made during the experiments, namely the vanishing impact of crosslinking for especially high specific currents. As discussed above, the microstructure of the sample with XG-PTMA-CMK-8 had a less restricting pore phase regarding shortest transportation paths, thereby having a positive effect on the rate capability of this sample. However, the current flow is of course still influenced by the microstructure. The impact of these restrictions increased with an increasing specific current. For very large specific currents, it is reasonable to assume that the whole pore network becomes a limiting factor to charge transfer, such that the differences between both samples become negligibly small, thereby negating the advantage of crosslinking.

## 5. Conclusions

In the present paper, we were able to show the positive effects of crosslinking PTMA on the electrochemical performance. For this purpose, two electrodes, one with ordinary PTMA and one with crosslinked PTMA, were compared based on their electrochemical characteristics and the 3D morphology of their microstructures. These electrodes were experimentally investigated by means of SEM, EDX mapping, and galvanostatic cycling. Moreover, statistical image analysis was performed to quantify the electrode microstructures based on 3D image data, which were obtained by XT. Compared to the spatial distribution of the non-crosslinked PTMA in the G-PTMA-CMK-8 electrode, the material in the electrode based on the crosslinked PTMA aggregates to larger clusters and larger pores emerge across the coating during the drying process. This leads to a morphology of the XG-PTMA-CMK-8 sample closely resembling the morphology of the conductive additive CMK-8. By resolving the 3D morphology using XT, we were additionally able to show that the large number of agglomerates in the XG-PTMA-CMK-8 sample leads to a morphology, where the lengths of shortest transportation paths through the pore space are significantly smaller compared to those of G-PTMA-CMK-8.

Furthermore, based on the galvanostatic cycling of the samples we were able to show that the electrochemical properties of the sample with crosslinking are clearly preferable, with a larger capacity, larger capacity retention after 100 cycles and higher rate capability. These improvements are presumably amplified by the microstructure differences, since the shorter transportation paths through the pore space and the larger interface between electrolyte and solid phase facilitate a homogenous dis-/charging of the electrodes. These findings lead to a better understanding of the positive influence of crosslinking the PTMA, and, more generally, provide a first impression on how the electrochemical properties of such polymer electrodes are influenced by the underlying microstructure.

**Acknowledgements**

All authors are grateful to the German Research Foundation (DFG) for funding their research projects within the framework of SPP 2248 ‘‘Polymer-based batteries’’ (BR 5752/4-1, MA 5039/7-1, SCHM 997/39-1). The work of M.N. was funded by the German Research Foundation (DFG) under Project ID 390874152 (POLiS Cluster of Excellence). D.B. and J.A. would additionally like to acknowledge the basic funding provided by the Helmholtz association. M.A. acknowledges funding by the Graduate & Professional Training Center Ulm. Besides, this study contributes to the research performed within CELEST (Center for Electrochemical Energy Storage Ulm-Karlsruhe).

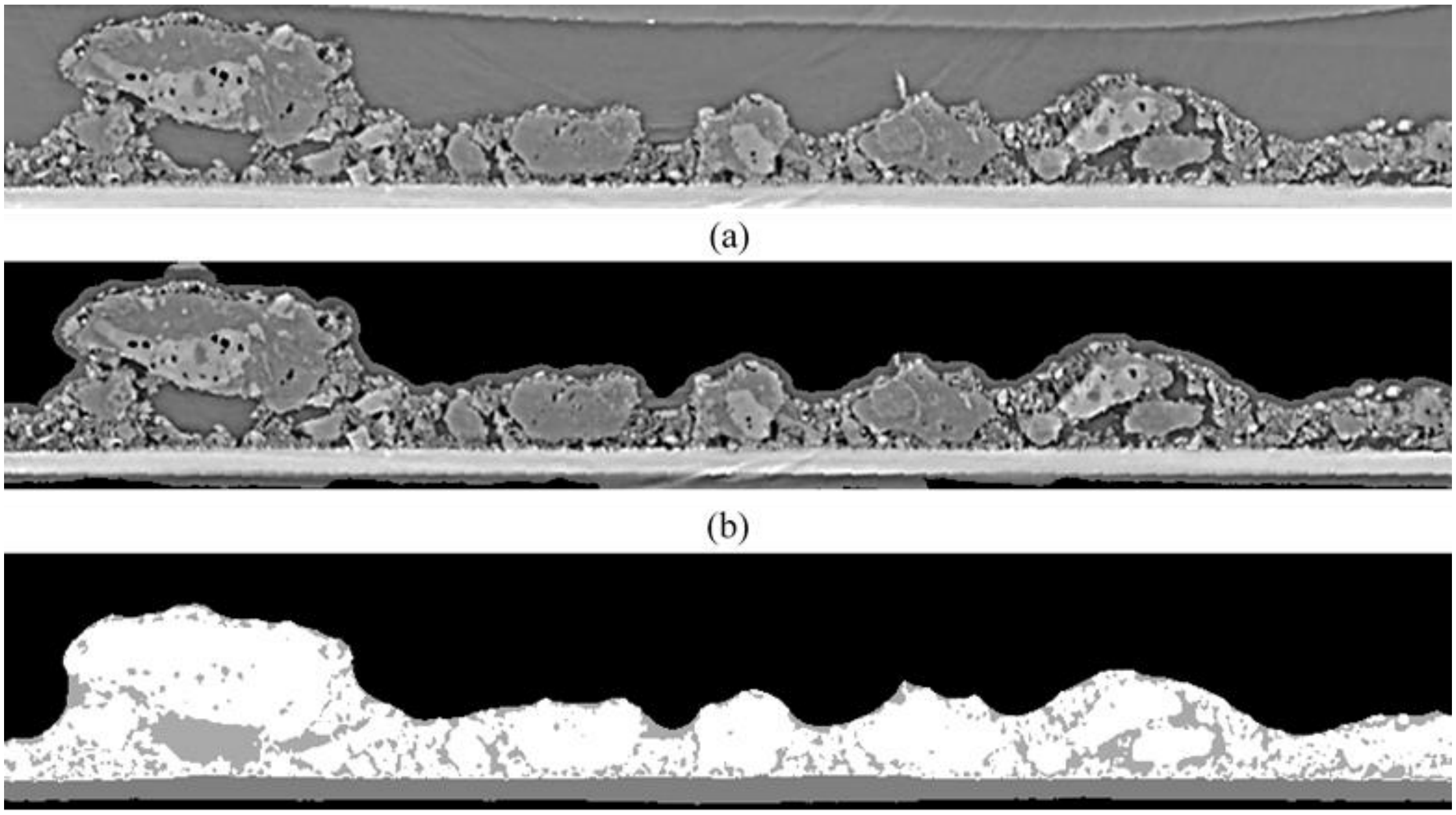
(a)
(b)
(c)

**Figure 1.** Exemplary slices of the sample with G-PTMA-CMK8 from the XT image data (a), the mask before phase segmentation (b) and the final segmentation (c).

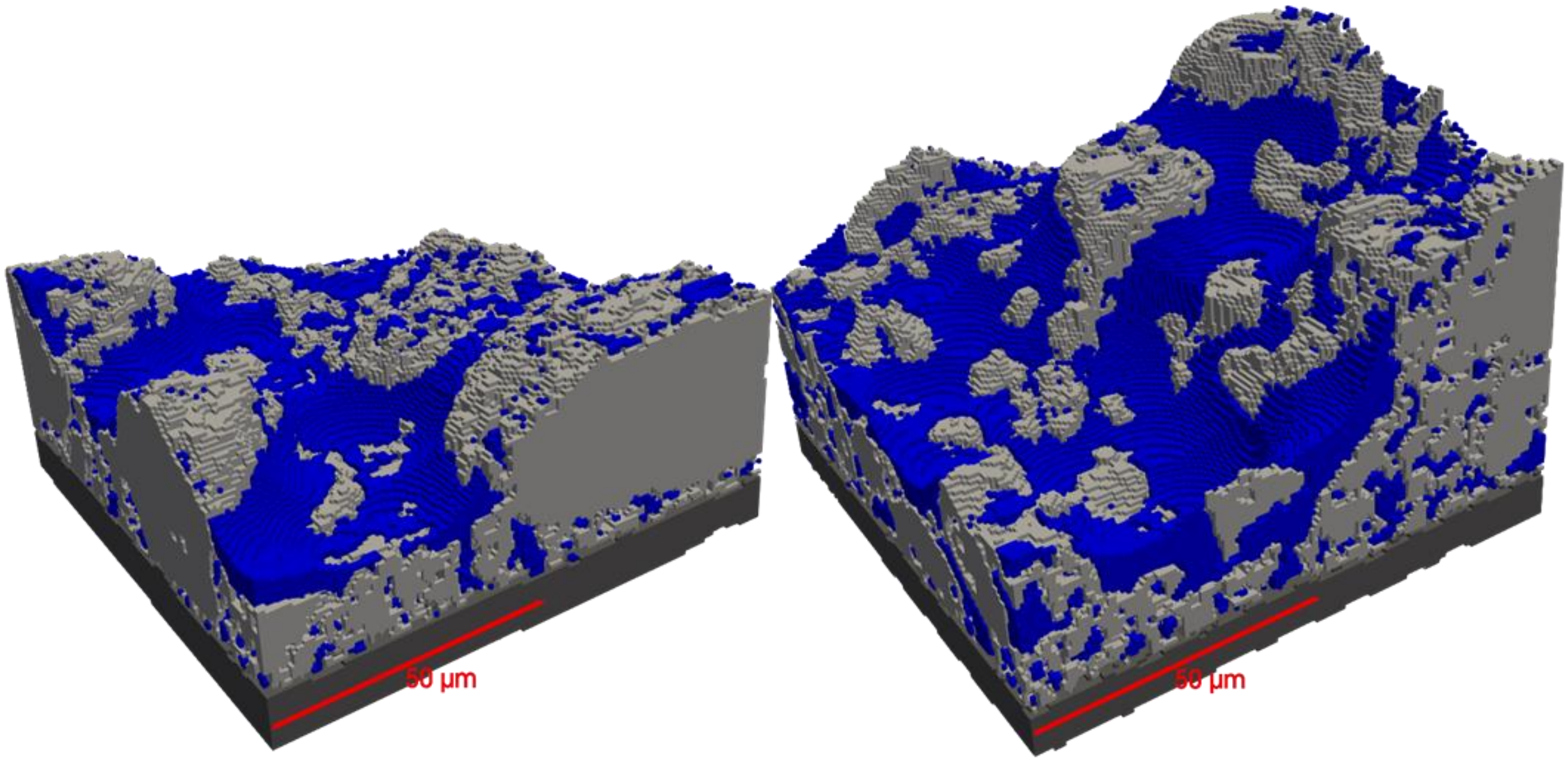

**Figure 2**. 3D rendering of segmented image data for the electrode based on G-PTMA-CMK-8 (left) and XG-PTMA-CMK-8 (right).

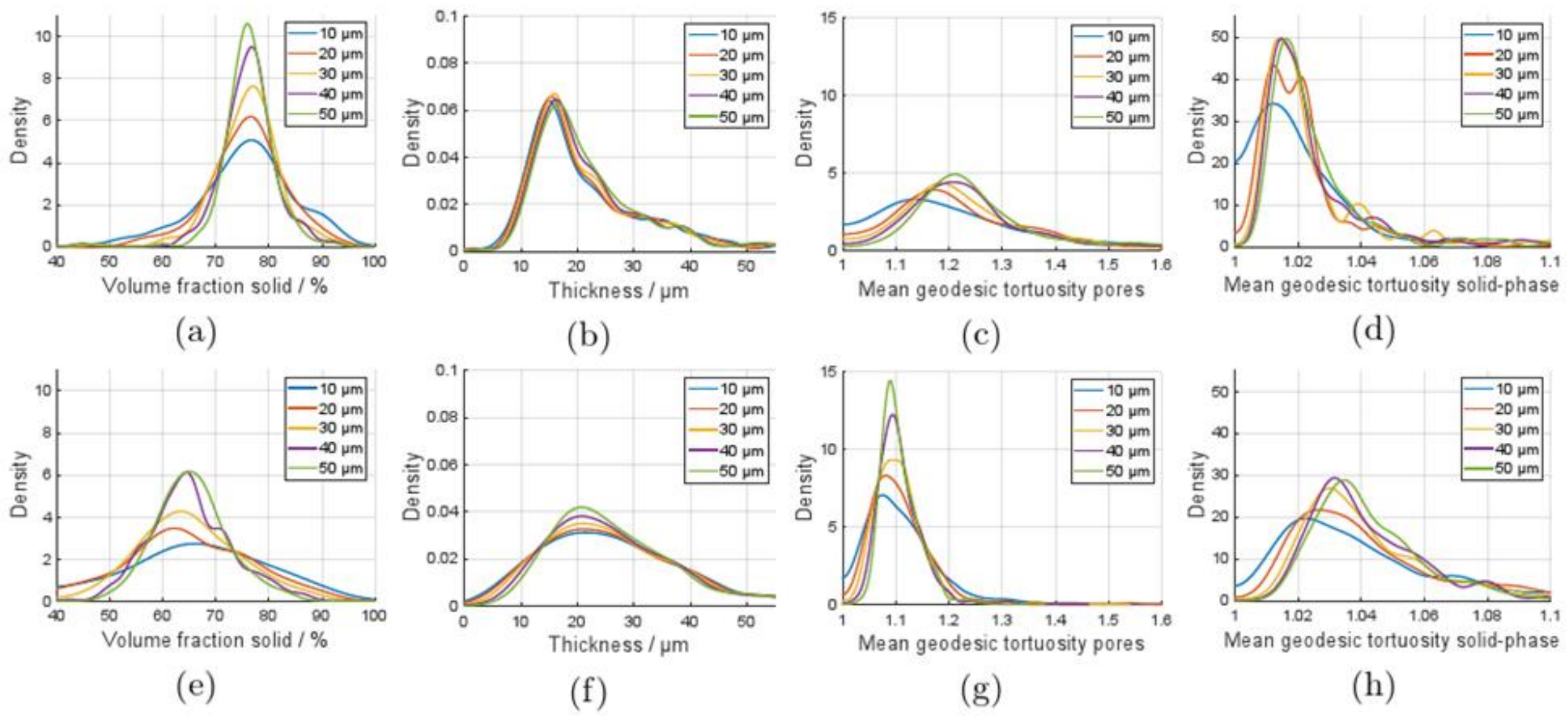

**Figure 3.** Univariate distributions of local microstructure descriptors (volume fraction, thickness, mean geodesic tortuosity pores, and mean geodesic tortuosity solid phase) of the G-PTMA-CMK8 electrode (top row) and the XG-PTMA-CMK8 electrode (bottom row) computed from cutouts with increasing window sizes.

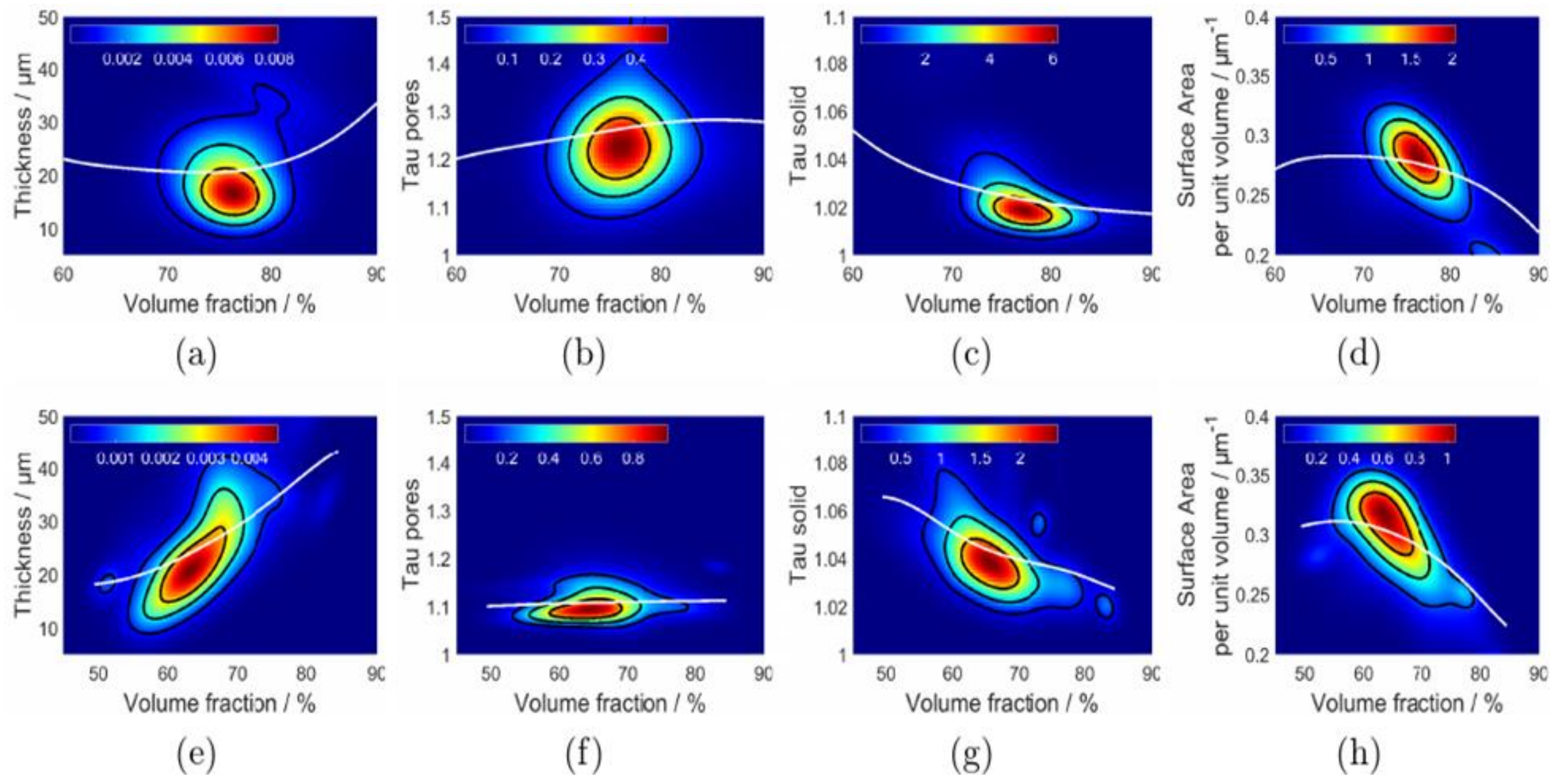

**Figure 4.** Bivariate distributions of pairs of local microstructure descriptors (thickness, τ pores, τ solid phase, and surface area) of G-PTMA-CMK8 (top row) and XG-PTMA-CMK8 (bottom row) computed from cutouts with a size of 50 µm. The bivariate probability density functions are visualized as heat maps, where the white line represents the (conditional) mean value for given local volume fractions of the polymer phase. The black contour lines are the 25%-, 50%-, and 75%-quantiles, respectively.

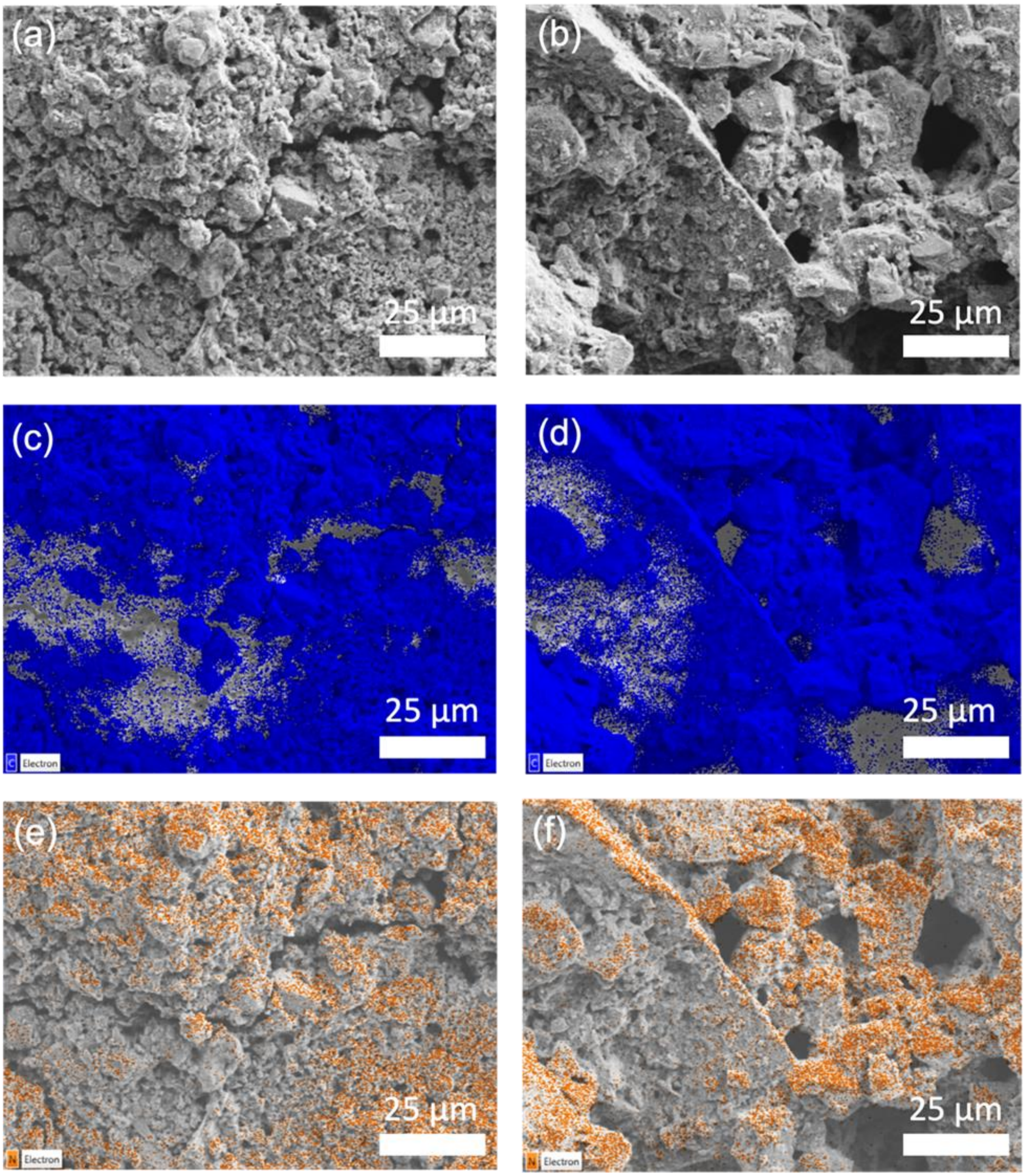

**Figure 5.** SEM micrographs of G-PTMA-CMK-8 (a) and XG-PTMA-CMK-8 (b) electrodes, and the corresponding EDX mappings of carbon (in blue) and nitrogen (in orange) for G-PTMA-CMK-8 (c,e) and XG-PTMA-CMK-8 (d,f).

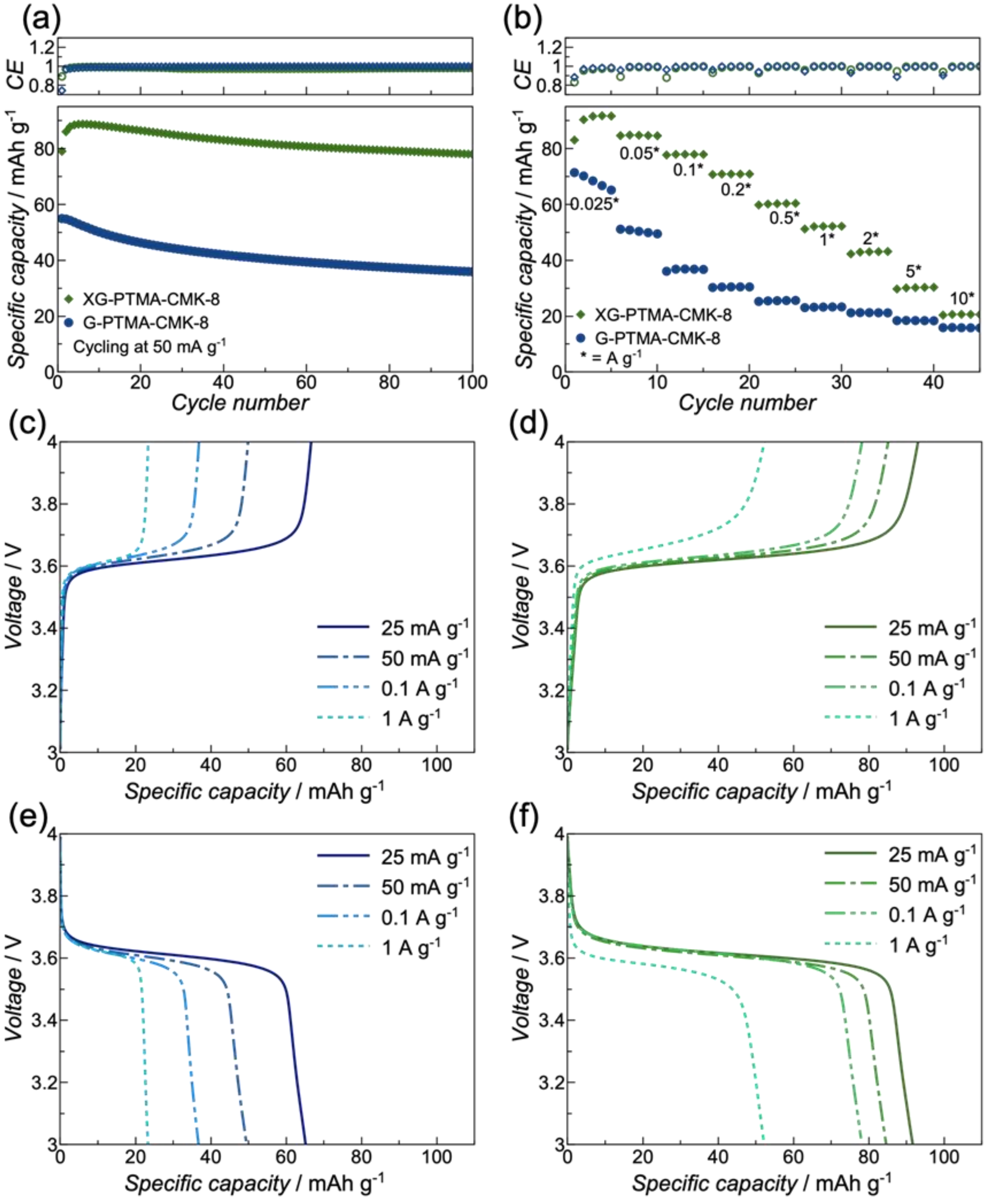

**Figure 6.** Galvanostatic cycling of G-PTMA-CMK-8 (in blue) and XG-PTMA-CMK-8 electrodes (in green): Comparison of the constant current cycling at 50 mA $g^{-1}$(a); (rate capability test at varying specific currents from 0.025 to 10 A $g^{-1}$(b); exemplary charge (c,d) and discharge profiles (e,f) of G-PTMA-CMK-8 (c,e) and XG-PTMA-CMK-8 (d,f) electrodes at specific currents ranging from 25 mA $g^{-1}$ to 1 A $g^{-1}$.

# Supporting Information

## Unveiling the Impact of Crosslinking Redox-Active Polymers on Their Electrochemical Behavior by 3D Imaging and Statistical Microstructure Analysis

Marten Ademmer [a,§], Po-Hua Su [b,c,§], Lukas Dodell [a], Jakob Asenbauer [b,c], Markus Osenberg [d], André Hilger [d], Jeng-Kuei Chang [e,f], Ingo Manke [d], Matthias Neumann [a,*], Volker Schmidt [a], Dominic Bresser [b,c,*]

[a] *Institute of Stochastics, Ulm University, 89069 Ulm, Germany*

[b] *Helmholtz Institute Ulm (HIU), 89081 Ulm, Germany*

[c] *Karlsruhe Institute of Technology (KIT), P.O. Box 3640, 76021 Karlsruhe, Germany*

[d] *Institute of Applied Materials, Helmholtz-Zentrum für Materialien und Energie, 14109 Berlin, Germany*

[e] *Department of Materials Science and Engineering, National Yang Ming Chiao Tung University, Hsinchu 30010, Taiwan*

[f] *Department of Chemical Engineering, Chung Yuan Christian University, Taoyuan 32023, Taiwan*

***Corresponding authors:** dominic.bresser@kit.edu; matthias.neumann@uni-ulm.de

[§]These authors contributed equally.

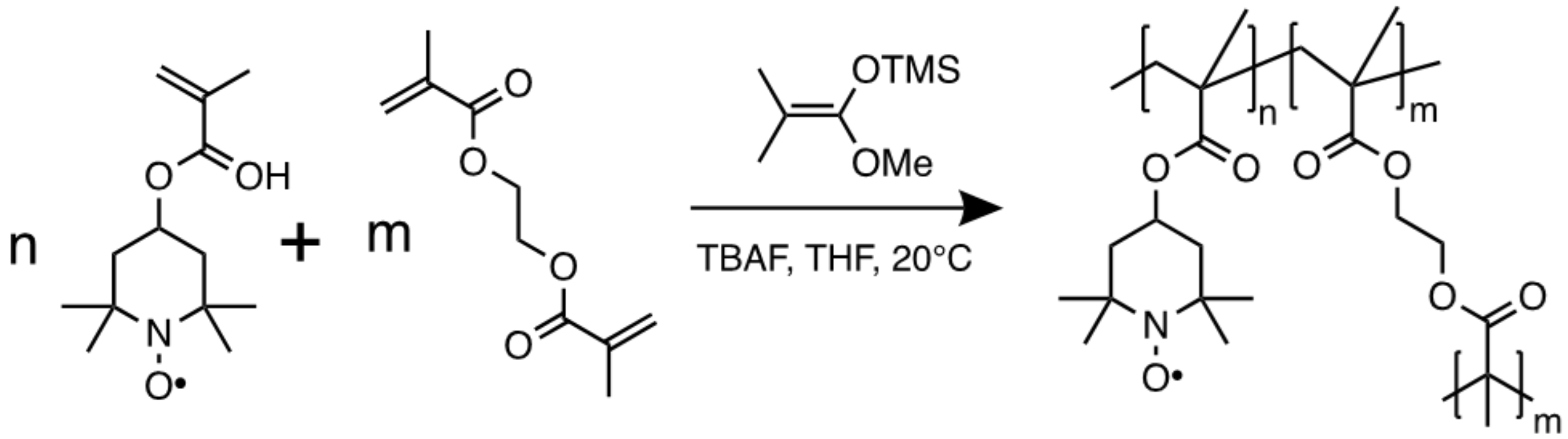

**Figure S1.** Synthesis scheme for yielding (crosslinked) PTMA with m = 0 for G-PTMA and m $\neq$ 0 for crosslinked XG-PTMA.

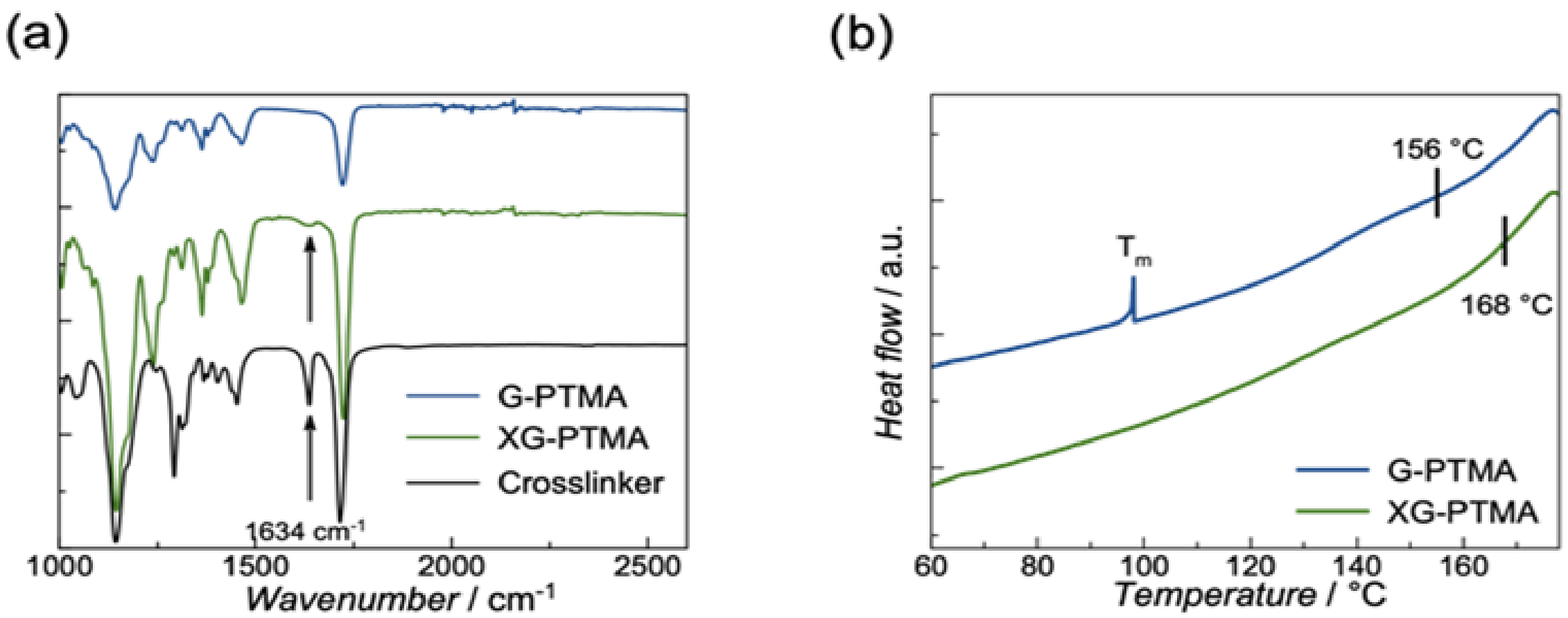

**Figure S2.** (a) FTIR spectra of G-PTMA (blue), XG-PTMA (green) and the cross-linker, ethylene glycol dimethacrylate (black). (b) DSC data recorded for (X)G-PTMA, conducted with a scan rate of 5 °C min$^{-1}$.

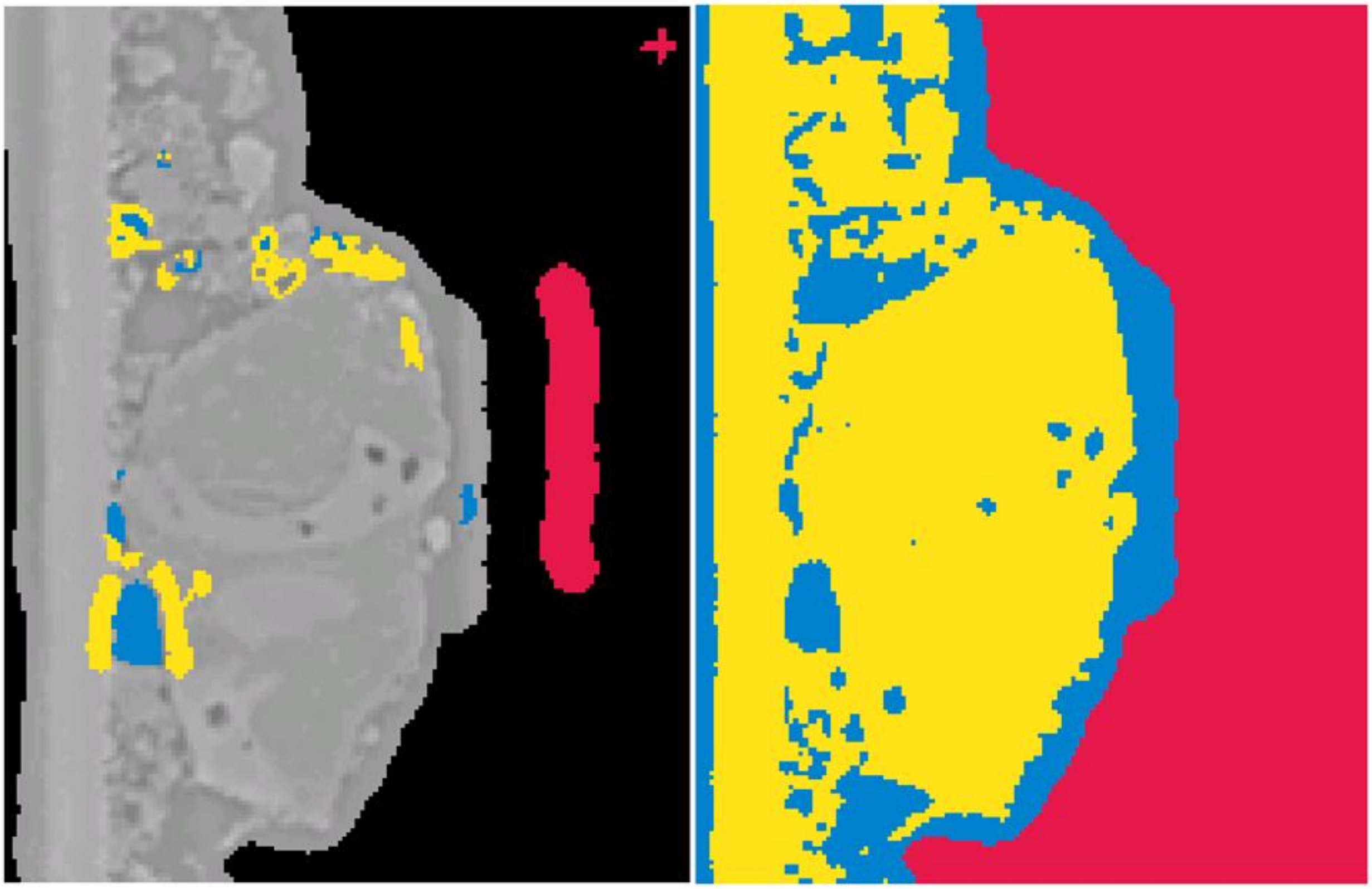

**Figure S3.** Example cutout of the hand-labeled training data of the sample with G-PTMA-CMK-8 on the left and the same cutout labeled by Ilastik on the right.

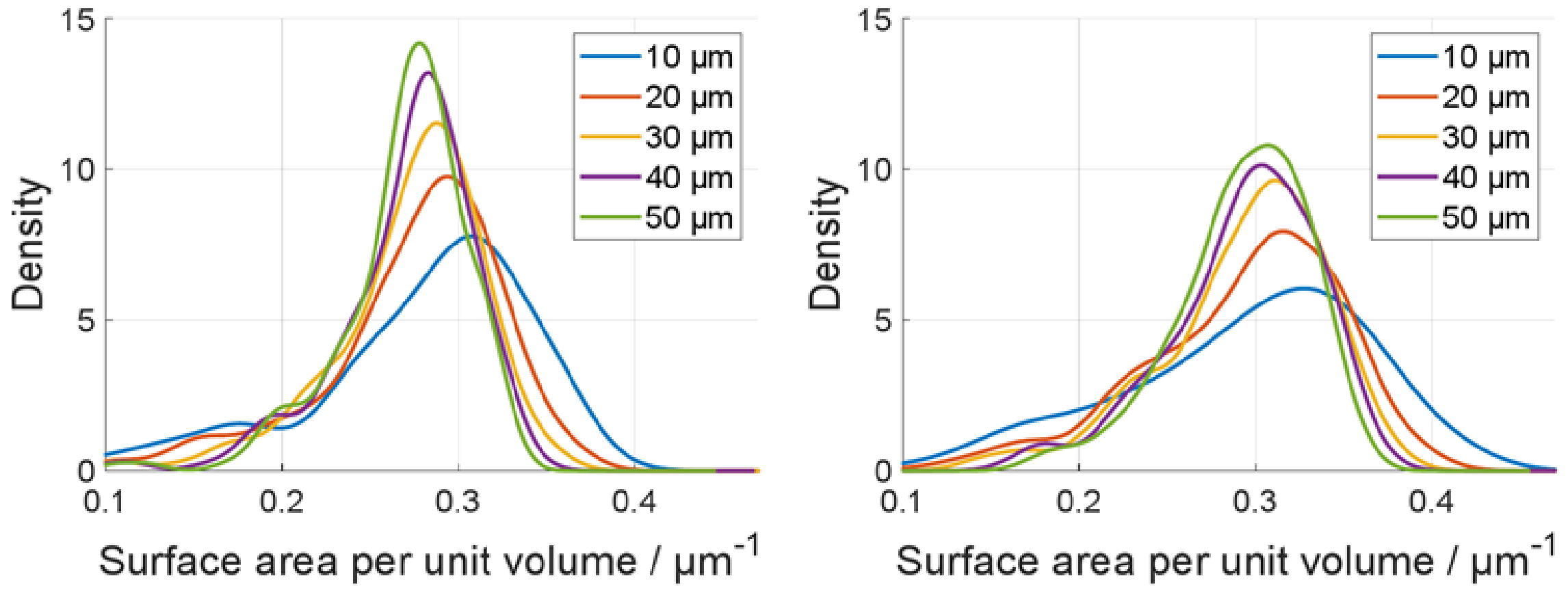

**Figure S4.** Distribution of local surface area per unit volume of G-PTMA-CMK-8 (left) and XG-PTMA-CMK-8 (right)

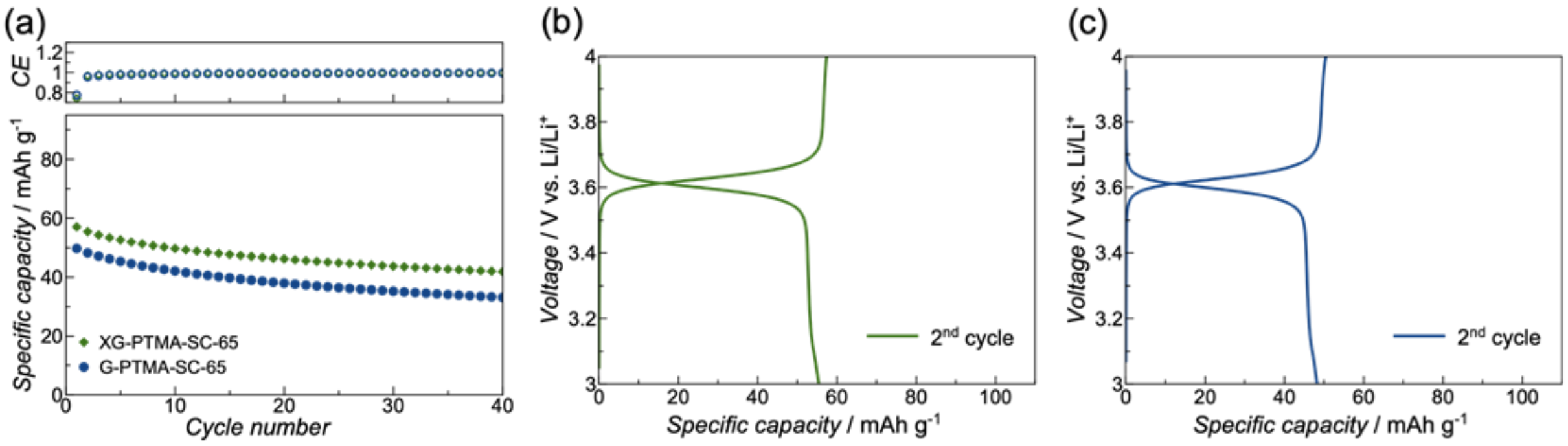

**Figure S5.** (a) Galvanostatic cycling of XG-PTMA-SC65 (green) and G-PTMA-SC65 electrodes (blue) at a constant specific current of 50 mA $g^{-1}$. The dis-/charge profile of the $2^{nd}$ cycle is displayed in panel (b) for XG-PTMA-SC65 and panel (c) for G-PTMA-SC65.

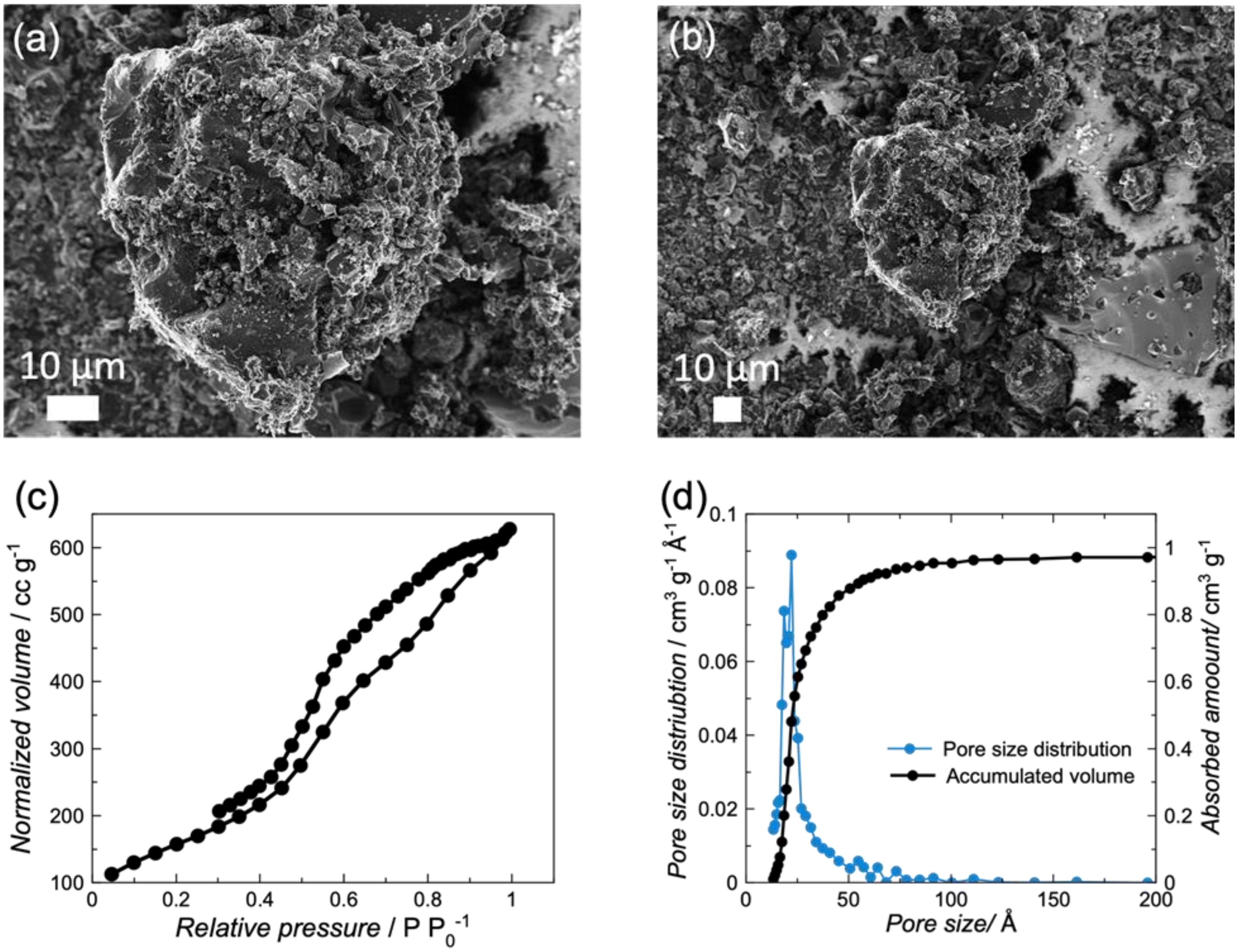

**Figure S6.** Basic physicochemical characterization of CMK-8 via (a,b) SEM at two different magnifications, (c) BET, and (d) BJH analysis.

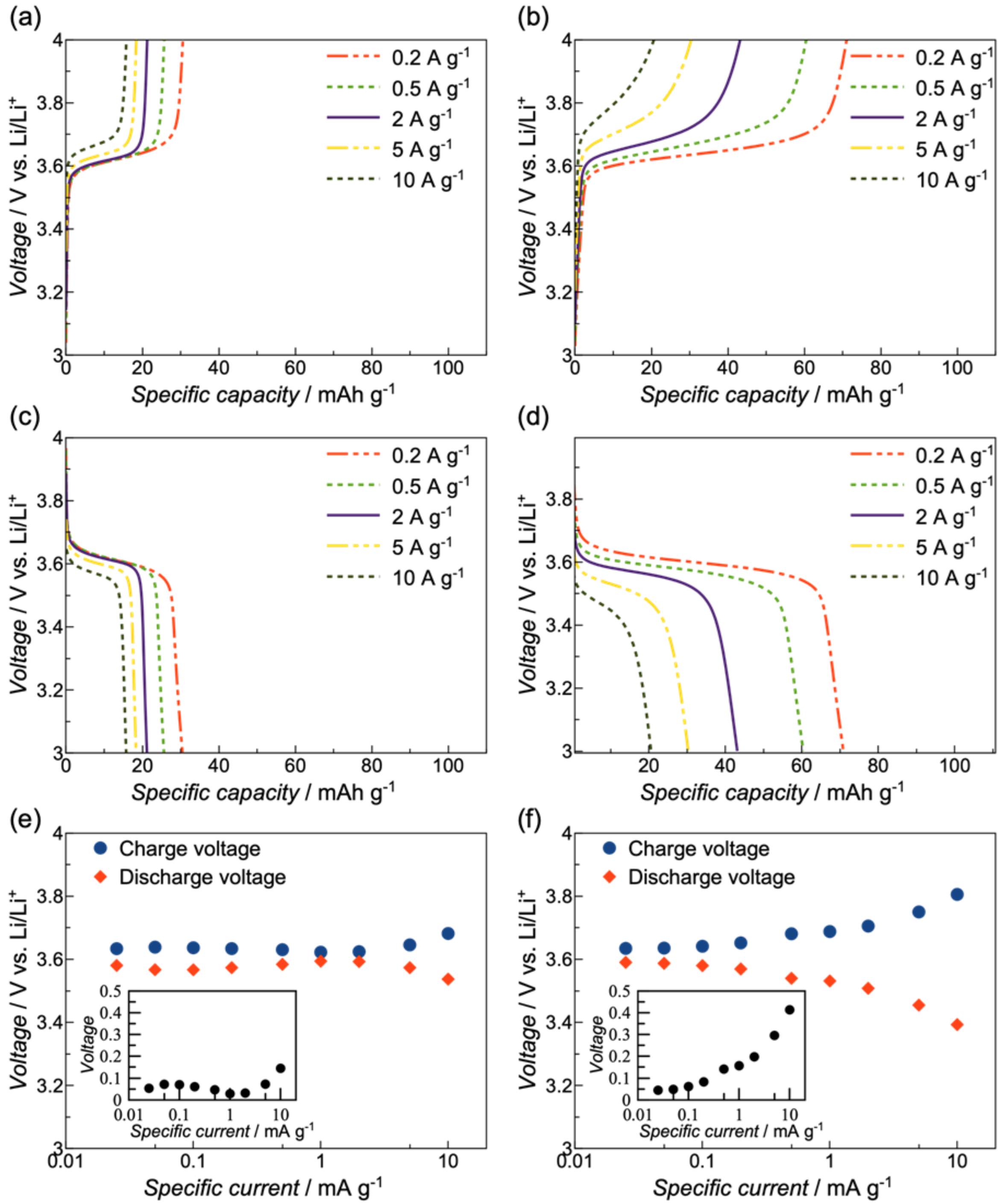

**Figure S7.** (a,b) Charge and (c,d) discharge profiles of galvanostatically cycled (a,c) G-PTMA-CMK-8 and (b,d) XG-PTMA-CMK-8 electrodes at specific currents varying from 0.2 to 10 A $g^{-1}$. Plots of the average charge and discharge voltage vs. the applied specific current (logarithmic) for (e) G-PTMA-CMK-8 and (f) XG-PTMA-CMK-8.